\documentclass[showpacs,prb,twocolumn]{revtex4}

\usepackage{graphicx}
\usepackage{color}\usepackage{enumerate}
\usepackage{amsmath}\usepackage{amssymb}\usepackage{stmaryrd}
\usepackage{multirow}
\usepackage{float}
\usepackage{longtable,booktabs}
\newcommand \beq {\begin{equation}}
\newcommand \eeq {\end{equation}}
\newcommand \beqa {\begin{eqnarray}}
\newcommand \eeqa {\end{eqnarray}}
\newcommand \dg {\dagger}

\newcommand \ran {\rangle}
\newcommand \lan {\langle}

\newcommand \mb {\mathbf}

\begin{document}

\title{Interaction Enabled Topological Crystalline Phases}

\author{Matthew F. Lapa, Jeffrey C. Y. Teo  \& Taylor L. Hughes}
\affiliation{ Department of Physics, Institute for Condensed Matter Theory, University of Illinois, 1110 W. Green St., Urbana IL 61801-3080, U.S.A.}

\begin{abstract}
In this article we provide a general mechanism for generating interaction-enabled fermionic topological phases. We illustrate the mechanism with crystalline symmetry-protected topological phases in 1D and 2D. These non-trivial phases require interactions for their existence and, in the cases we consider, the free-fermion classification yields only a trivial phase. Similar to the interpretation of the Kitaev Majorana wire as a mean-field p-wave superconductor Hamiltonian arising from an interacting model with quartic interactions, we show that our systems can be interpreted as ``mean-field" charge-$4e$ superconductors arising, e.g., from an interacting model with eight-body interactions, or through another physical mechanism. The quartet superconducting nature allows for the teleportation of full Cooper pairs, and in 2D for interesting semiclassical crystalline defects with non-Abelian anyon boundstates. 
\end{abstract}

%\pacs{}
\maketitle
%\tableofcontents

Symmetry protected topological phases (SPTs) have sprung to the forefront of condensed matter physics. The impetus for such 
an explosion of interest began with the theoretical prediction and experimental discovery of 2D and 3D topological insulators protected by time-reversal symmetry\cite{HasanKane} and has carried on through the classification of all weakly-interacting fermionic topological phases protected by discrete (anti-unitary) symmetries\cite{kitaev,schnyder,QHZ2008}. From here the field has now spread to encompass topological crystalline phases protected by spatial symmetries\cite{teo2008,fu2011topological,hughes2011inversion,turner2012, fang2012bulk,teohughes1,slager2012, Benalcazar2013,Morimoto2013,RyuReflection2013,fang2013entanglement,HughesYaoQi13,Sato2013,Kane2013Mirror,jadaun2013}, and bosonic counterparts of the fermionic phases\cite{Wen2011classification,Wen2011complete,levin2012,VishwanathSenthil2013,KaneFisher2013}, which recently culminated into a classification of some interacting SPTs\cite{Vishwanath2012CS,Wen2013,Senthil2014}. In this article we develop a mechanism for interaction-enabled fermionic crystalline SPTs and provide explicit 1D and 2D models that realize the putative strongly-interacting topological phases\footnote{We will usually not distinguish between SPTs protected by local symmetries, and crystalline SPTs protected by possibly non-local point-group symmetries}. We show that, without interactions, the symmetry classes we consider have no non-trivial topological phases, yet with interactions a non-trivial SPT exists. The models we consider are essentially related to charge-$4e$ superconductors and have no mean-field (free-fermion) description. We also discuss the strongly interacting topological phase transition between the trivial and interaction-enabled topological phase.

We begin by recounting the theory of one of the simplest topological phases, the Kitaev p-wave wire with an additional time-reversal symmetry $T$ ($T^2=+1$)\cite{kitaev2001}. This model belongs to the BDI Altland-Zirnbauer class\cite{AZ, schnyder} and is classified, in the non-interacting limit, by an integer winding number $\nu.$ An example of a model in this class realizing the $\nu=1$ phase is shown in Figure 1a. The model includes one spinless fermion per unit cell $\psi_n$, which is conventionally split into two Majorana fermions $\psi_n= \frac{1}{2}(a_n+i b_n)$ satisfying $a_n=a_{n}^{\dagger},\; b_{n}=b_{n}^{\dagger}$ and $Ta_nT^{-1}=a_n,\; Tb_nT^{-1}=-b_n.$ The last constraint is inherited from the action of time-reversal symmetry on $\psi_n,$ i.e., $T\psi_n T^{-1}=\psi_n.$ For each line in Figure 1a that connects Majorana fermions, a tunneling term of the form $ib_{n}a_{n+1}$ appears in the Hamiltonian. These terms are Hermitian and time-reversal invariant. The coupling between the Majorana modes opens an energy gap in the bulk, but the system harbors low-energy Majorana zero-modes on the ends of an open chain. The bulk-boundary correspondence dictates that the number of boundary zero modes of $a$-type minus the number of $b$-type on a single end is $\vert \nu\vert.$ In Figure 1a one can see the minimal configuration for $\nu=1$ with one $a$-mode on the left end and a $b$-mode on the right end. Larger values of $\nu$ are topologically equivalent to multiple copies of the $\nu=1$ case and exhibit $\vert \nu\vert $ stable Majorana modes of an identical type on a single end. Negative values of $\nu$ correspond to chains with unpaired $a$-modes ($b$-modes) on the right (left).

With this in mind we wish to consider the possibility of a crystalline topological superconductor in this class by requiring inversion symmetry $R$ with $R^2=1$ and $[R, T]=0$, which is natural for spinless (or spin-polarized) fermions.  Unfortunately the classification is not very interesting. In class BDI, the fact that $R$ and $T$ commute means that $R$ does not act within a unit cell to interchange $a$ and $b$ type modes. Thus, we can see from Figure 1a that acting with $R$ just flips the chain from left to right and will convert an $a$-type end to a $b$-type end which implies that $R\colon \nu\to -\nu.$ Thus, when the symmetry is enforced we must have $\nu=-\nu$, but the only solution is $\nu=0$ since $\nu$ is an integer. Hence, there are no free-fermion SPTs for the BDI class with inversion symmetry, nor any weakly-interacting SPTs in this symmetry class that can be adiabatically connected to the non-interacting limit. 

This is not the end of the story. It has been explicitly shown that the classification of the vanilla BDI class with interactions is deformed from its non-interacting limit\cite{FK2010,FK2011,TPB2011,Wen2011complete}. In a seminal paper\cite{FK2010} Fidkowski and Kitaev showed that eight copies of the $\nu=1$  chain (i.e. $\nu=8$) can be adiabatically deformed to $\nu=0$ by passing through a gapped, interacting phase that preserves all of the required symmetries. Hence, the integer classification is reduced to $\mathbb{Z}_8.$ Remarkably, if we add inversion symmetry, the constraint $\nu=-\nu$ now has a non-trivial solution! Since $\nu$ is defined modulo eight, the solution $\nu=4\equiv-4\mod 8$ indicates the existence of a non-trivial crystalline SPT that \emph{requires} strong interactions for its existence.

This mechanism for, what we call, an interaction-enabled SPT is quite general. Given any topological integer property $\nu,$ and a symmetry $R$ under which $\nu$ transforms non-trivially, then the constraint $\nu=R\nu$ has no non-trivial solutions, i.e. $\nu\equiv 0$ is the only solution. The property $\nu$ could be a scalar, vector, tensor etc., but for now let us focus on the scalar case where $\nu$ can only transform non-trivially to $-\nu.$ By including interactions, or other deformations of the Hamiltonian, the integer classification of $\nu$ could be reduced to a cyclic group $\mathbb{Z}_n.$ If $n$ is even, then $\nu=n/2$ is a non-trivial solution, and the classification of the interacting system with $R$ symmetry is $\mathbb{Z}_2$ valued where $\nu=0\mod n$ and $\nu=n/2\mod n$ are the trivial and non-trivial values respectively. In the remainder of the article we will construct and discuss the properties of 1D and 2D models which have interaction-enabled topological phases. 

Let us begin with our paradigm case of class BDI with inversion symmetry. Although we will not provide an explicit proof that it cannot be done, let us illustrate the complication with generating the non-trivial $\nu=4$ state from a free-fermion (quadratic in fermion operators) Hamiltonian. To preserve $T$ the only allowed quadratic terms must couple an $a$-type and a $b$-type fermion, i.e., terms of the form $ia_{n}b_{m}.$ To preserve inversion symmetry, each end of the chain must have the same number, and type, of low-energy Majorana modes. Thus, beginning with the ends of a topological chain and working backward to form the gapped bulk, one can show that there must always be gapless states in the bulk if we use free fermions. One explicit example is shown in Figure 1b, where one eventually reaches a place on the chain where fermions of the same type must be coupled to open a gap, but this is forbidden by time-reversal. While we have only shown a specific example, it is generally true that one cannot create this topological phase from purely free Hamiltonians.

This failure, however, immediately gives the key to the correct construction. We see that what is needed is a perturbation that can open a gap by coupling eight Majorana fermions of the same type in an inversion and time-reversal symmetric way. Fortunately the Fidkowski-Kitaev interaction is exactly what is needed (see Appendix A). In fact, if we couple the eight bulk Majorana zero-modes in Figure 1b with the FK interaction then we will have a fully gapped $\nu=4$ topological phase that preserves $T$ and $R$ and intrinsically requires interactions for its existence. This inhomogeneous chain may seem a little strange, so instead let us consider a translationally invariant model which we call the Fidkowski-Kitaev (FK)-chain. Instead of coupling Majorana zero-modes with quadratic tunneling terms, we couple the modes with the quartic FK interaction in a ``dimerized" pattern which we now discuss.

\begin{figure}[t]
  \centering
    \includegraphics[width= .5\textwidth]{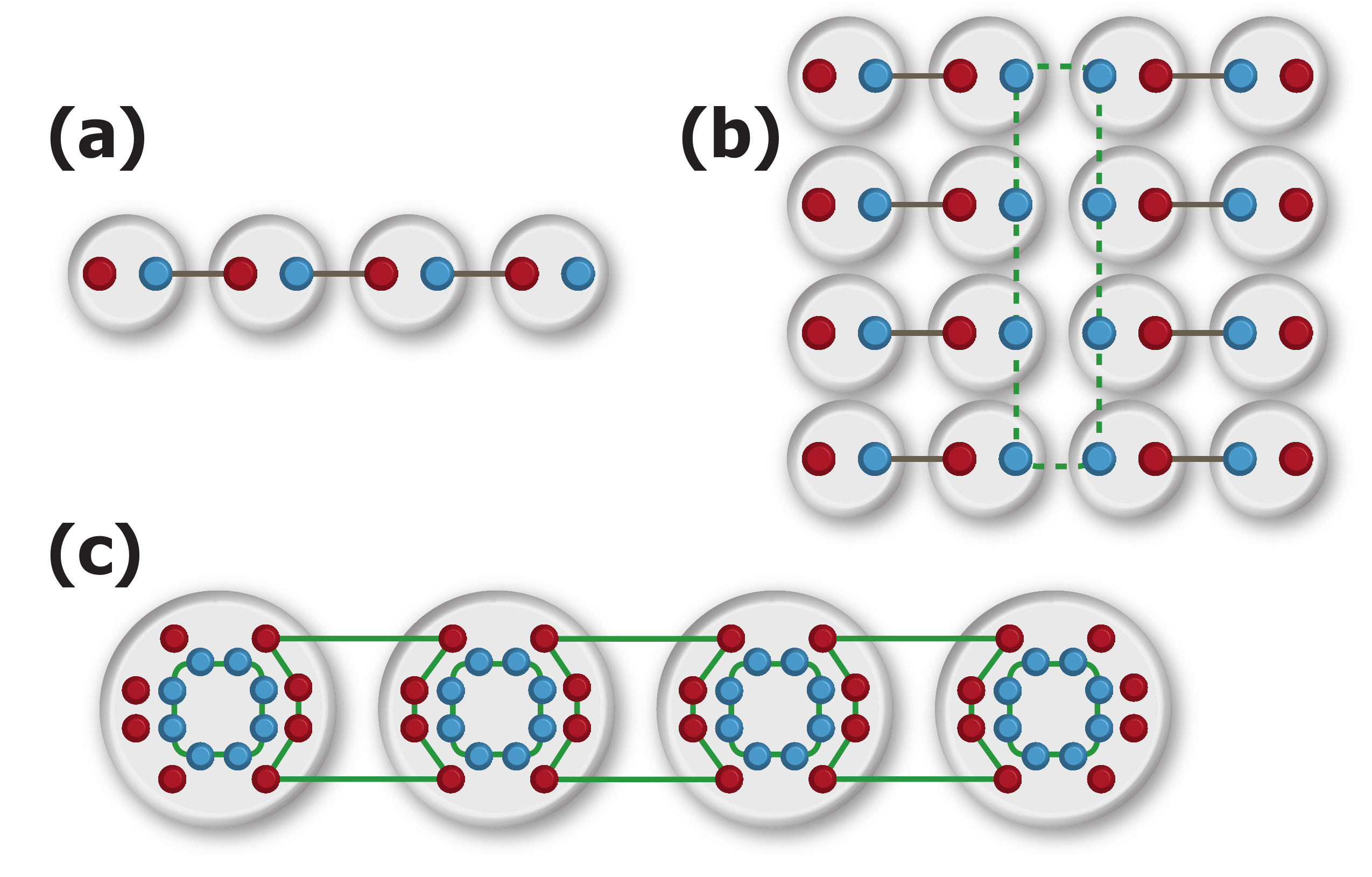} 
\vskip 10pt
 \caption{\emph{Topological Superconductors in 1D:} (a) The Kitaev p-wave wire with time-reversal symmetry $T (T^2=+1)$ . Each large circle represents one unit cell, which contains one
complex fermion. The complex fermion is then split up into a-type (red) and b-type (blue) Majorana fermions. In the topological
phase the b-type Majorana fermion from the nth unit cell is coupled to the a-type Majorana fermion in the n+1 th unit cell 
with a quadratic hopping term $ia_{n+1}b_{n}$ (represented here by a grey line). (b) An attempt to construct an inversion and time-reversal symmetric topological phase using a free-fermion model. This system must have
 four unpaired Majorana fermions of the same type on each end of the wire and this will always lead to gapless states in the bulk. The gapless states cannot be gapped out using a 
quadratic interaction without breaking the time-reversal symmetry and thus we need the Fidkowski-Kitaev interaction (green dotted line) to open a bulk gap. (c) The Fidkowski-Kitaev (FK)-chain model. Each unit
cell (large white circle) contains eight complex fermions, which can be split into eight a-type and eight b-type Majorana fermions.
We couple the eight b-type Majorana fermions in each unit cell with the quartic FK interaction (represented by the green lines)
and we also couple four of the a-type Majorana fermions in the right side of a unit cell with the four a-type Majorana fermions
on the left side of the adjacent unit cell using the FK interaction.}
\label{fig:1Dmodel}
\end{figure}

The FK-chain model consists of eight complex fermions $\psi^J_n$ per unit cell, where $J= 1,\dots,8$ is a flavor index. Each
complex fermion can be split into two Majorana modes $\psi^J_n = \frac{1}{2}(a^J_n + ib^J_n)$ as before. 
In the FK-chain we couple
the eight b-type Majorana modes $b^J_n$ \emph{within} each unit cell using the FK-interaction, and we also couple the four a-type
Majorana modes $a^J_n$, $J= 5,\dots,8$ in unit cell $n$ with the four Majorana modes $a^J_{n+1}$, $J=1,\dots,4$ in unit
cell $n+1$ using the FK-interaction, as shown in Figure 1c. The resulting Hamiltonian (which is explicitly written in Appendix A )
has the following very important properties: 1) it is time-reversal and inversion symmetric, 2) it has a unique gapped ground
state on a periodic chain, and 3)  each end of an open chain harbors two effective spin 1/2 degrees of freedom (4 Majorana modes), 
and the time-reversal operator acts projectively as $T^2 = -1$ on each of these spin 1/2 degrees of freedom. This is the non-trivial 1D crystalline topological phase with $\nu=4.$

The pair of boundary spin-1/2's on each end are composed of four $a$-type Majorana fermions and are unstable in the presence of the most general time-reversal invariant perturbations.  On the left boundary, the four Majorana fermions $a^1, a^2, a^3$ and $a^4$ will be unpaired. 
Since these four Majorana modes transform in the same way under the action of $T$, quadratic terms of the form $ia^Ia^J$ are 
forbidden; the only Hermitian and time-reversal invariant
term we can add is of the form $H_{bdy} = \lambda a^1a^2a^3a^4$. 
This term is essentially a symmetrized Hubbard-like interaction, which can be seen by defining the new complex fermions 
$\chi_1 = \frac{1}{2}(a^1 + ia^2)$, $\chi_2 = \frac{1}{2}(a^3 + ia^4)$. In terms of the $\chi_i$ we have
\begin{equation}
	H_{bdy} = -\lambda(2\chi^{\dg}_1\chi_1 - 1)(2\chi^{\dg}_2\chi_2 - 1)=-\lambda(-1)^{F_{\chi_1,\chi_2}}\  \label{eq:Hbdy}
\end{equation} where $(-1)^{F_{\chi_1,\chi_2}}$ is the local fermion parity at the boundary. This local Hamiltonian has two degenerate ground states $\vert 0\rangle_g, \vert 1\rangle_g$ and two degenerate excited states $\vert 0\rangle_e, \vert 1\rangle_e.$ If $\lambda >0$ ($\lambda <0$) the ground states both have even (odd) fermion parity and vice-versa for the excited states. As shown in Appendix B, time-reversal acts non-trivially as $T_{bdy} = i\sigma^y K$ on both the ground and excited state subspaces independently.
 It follows immediately from Kramer's theorem that the remaining two-fold degeneracy of the boundary states is protected against 
arbitrary perturbations that do not break time-reversal symmetry.
Thus, even when the local fermion parity is locked by $H_{bdy},$ the low energy degrees of freedom on the edge still form a projective representation of the on-site time-reversal symmetry group\cite{Wen2011classification,Wen2011complete}, and the remaining degree of freedom in the lowest energy sector is a single spin-1/2. 

When the local fermion parity is fixed and does not fluctuate, the low energy properties of the FK chain are similar to the gapped (Haldane) phase of a spin-1 chain  protected by inversion and time-reversal symmetry\cite{haldane,AKLT}. Both systems
have a bulk gap and gapless spin-1/2 excitations at the boundary. In both models, $T$ acts as $T^2 = 1$ on the fundamental degrees of freedom
within each unit cell, but as $T^2 = -1$ on the fractionalized degrees of freedom at the ends of an open chain. However,
the two systems are not identical as the Hilbert space of the FK chain is necessarily larger than the Hilbert space of the
spin-1 chain due to the fermionic nature of the local degrees of freedom. It is only in the low energy subspace of the FK chain, with the interaction term $H_{bdy}$ included on each boundary unit cell to lock the local fermion parity, in which the similarity between the two systems becomes apparent.

Interestingly, the FK-chain only contains terms which are quartic
in fermion creation and annihilation operators, thus this system has no free fermion analogue. Indeed, the complete
two-particle Green function $\mathcal{G}(\omega,k)$ (i.e., the matrix of two-point functions with the regular time-ordered
Green functions on the diagonal blocks and the anomalous time-ordered Green functions on the off-diagonal blocks)
for this model vanishes at $\omega = 0$, which means that there is no 
Bogoliubov-de-Gennes (BdG) mean-field Hamiltonian that captures the physical properties of this system 
(recall that if $\mathcal{G}(0,k)\neq 0$ then we can construct a BdG Hamiltonian $H_{BdG}(k) \sim \mathcal{G}^{-1}(0,k)$ 
which defines a free
fermion system with topological properties identical to those of the interacting system).
The Hamiltonian for the FK chain contains
terms of the form $\psi^I_n \psi^J_n \psi^I_{n+1} \psi^J_{n+1} + \psi^{I,\dg}_n\psi^{J,\dg}_n\psi^{I,\dg}_{n+1}\psi^{J,\dg}_{n+1}$,  leading to a 
non-vanishing anomalous four-point function with momentum-dependence. Due to the integrability of the model, all 4-point correlation functions can be evaluated exactly, and in particular, charge conservation symmetry is  broken by $4e$ tetrads such as $\langle\psi^1_{k_1}\psi^2_{k_2}\psi^3_{k_3}\psi^4_{-k_1-k_2-k_3}\rangle\sim1+e^{ik_1+ik_2}.$ If one tunes away from the exactly solvable point (by for example adding quadratic tunneling or pairing terms) then one can extract an effective Hamiltonian from the inverse of the two-point function, however this Hamiltonian will be topologically trivial and will not contain the essential features of the topological phase which will still be contained in the four-point functions. We suspect a topological invariant could be constructed from these momentum-dependent 4-point correlation functions analogous to Ref.~\citenum{gurarie,gurarie1D}.

The existence of non-vanishing anomalous four-point 
functions for this model shows that the FK chain has broken charge-conjugation symmetry. For this model to arise microscopically we would expect this symmetry to be broken spontaneously via some ``mean-field" like state of an eight-body interacting Hamiltonian, or from a mechanism analogous to the charge-$4e$ superconductivity formed from a melted pair-density wave state in Ref. \onlinecite{BergFradkinKivelson}. Thus, similar to Kitaev's interpretation of the Majorana chain as a mean-field description of a spontaneously generated topological p-wave superconductor, our model can be essentially interpreted as a topological charge $4e$ superconductor.  We also note that a topological phase transition between a $\nu=4$ phase and a trivial charge-$4e$ superconductor with $\nu=0$ can be driven by turning on intra-cell FK couplings for the $a$-fermions and leaving the $b$-fermions unmodified. As discussed in the Methods section, the critical theory -- when the intra-cell FK interaction strength matches that of the inter-cell one -- can be described by a conformal field theory with central charge $c=1.$ It can be mapped into a non-chiral $u(1)_1$ boson theory with compactification radius $R=\sqrt{2}$ and carries an affine Kac-Moody $su(2)$ structure at level 1. 

%\cite{Afflecklecture}

%We now discuss the properties of a boundary site of the FK chain on a system with open boundary conditions. We mentioned
%above that the low energy states at one boundary of the FK chain organize into two effective spin 1/2 degrees of freedom
%(i.e. the boundary Hilbert space is a direct sum of two 2D spinor representations of $SU(2)$). We will show that 
%each boundary still retains one gapless spin 1/2 degree of freedom, even in 
%the presence of the most general time-reversal invariant perturbations.

%{\bf{TLH: Insert Cooper pair teleportation here and other measurement methods.}}
Unlike an ordinary BCS superconductor, Cooper pairs are finite energy excitations in a gapped $4e$ superconductor since they are not the fundamental bosons in the condensate. As a result, one would expect transport across a $4e$ superconductor sandwiched between normal superconducting leads should be dominated by four fermion Andreev reflection where a Cooper pair reflects off the $4e$ superconductor as a pair of holes and a $4e$ quartet propagates across the $4e$ superconductor. This process could be observed by shot noise in principle. While this is the case for a trivial charge $4e$ superconductor, our model also allows for another type of anomalous transport process.  Similar to a zero bias Josephson current across a topological BCS superconducting nanowire between normal metallic leads~\cite{law2009majorana, lutchyn2010majorana, oreg2010helical, mourik2012signatures}, we expect an anomalous zero bias differential conductance and a topologically enhanced double Andreev reflection across the topological $4e$ superconductor when normal superconducting leads are directly coupled to the edges. We can understand this as follows. Just as a single Majorana end state allows for single-electron teleportation by forcing the ground states with even and odd fermion parity to be degenerate\cite{Fu2010M}, the boundaries of the FK chain allow for Cooper pair teleportation since the ends force the ground states with an even and odd number of \emph{Cooper pairs} to be degenerate (which is not generic in a $4e$ superconductor). 

%This Cooper pair teleportation phenomena could be supported by the fourfold degenerate ground state of the boundary spin-$1/2$ degrees of freedom. Fermion pairs fusion channels will be shown in the Methods section.

 We now move on to consider some topological phases of two-dimensional superconductors in the BDI class, with translation and discrete rotation symmetries, i.e. topological crystalline superconductors (TCS)\cite{teo2008,fu2011topological,hughes2011inversion,turner2012, fang2012bulk,teohughes1,slager2012, Benalcazar2013,Morimoto2013,RyuReflection2013,fang2013entanglement,HughesYaoQi13,Sato2013,Kane2013Mirror,jadaun2013}. 
Generically these TCS's can carry a number of different non-trivial topological invariants, each of which is stable in the presence  of a certain subset of the symmetries of the model. 
Here we are only interested in the \emph{weak invariant}, an invariant which is stable in 
the presence of translation symmetry\cite{FuKaneWeak2007}. Heuristically, the weak invariant in 2D is a vector topological invariant generated by stacking 1D topological wires into 2D, and is thus necessarily anisotropic. The stacks of topological wires define a 2D lattice with reciprocal lattice vectors $\mb{b}_1$ and $\mb{b}_2$ and the weak invariant must take the form \beq
	\mb{G}_{\nu} = \frac{\nu_1}{2} \mb{b}_1 +\frac{\nu_2}{2} \mb{b}_2 \ ,
\eeq where $\nu_1$ and $\nu_2$ are integers for the BDI class since they arise from the 1D integer topological invariant.
As long as translation symmetry is protected, then two phases with different weak invariants cannot be 
adiabatically connected without closing the gap or breaking a symmetry. 

Just as for the 1D case, we want to require additional spatial symmetries. To be explicit let us choose $C_4$ rotation symmetry, which implies $\mb{b}_1=\tfrac{2\pi}{a}\hat{\mb{x}}$ and $\mb{b}_2=\tfrac{2\pi}{a}\hat{\mb{y}}$ 
(with lattice spacing $a$).  Just like the case of $\nu$ under inversion in 1D, $\mb{G}_{\nu}$ transforms non-trivially (i.e., as a vector) under $C_4$ symmetry. Enforcing the symmetry constrains the weak index to satisfy $G_{\nu}^x=G_{\nu}^{y}$ and $G_{\nu}^y=-G_{\nu}^x.$ The only values of the integers $\nu_1,\nu_2$ which satisfy this constraint are $\nu_1=\nu_2=0$\cite{Benalcazar2013}. However, just as above, if we allow for interactions then $\nu_1, \nu_2 \in \mathbb{Z}_8$ which means that $\nu_1=\nu_2=4$ is also a valid possibility, but one that requires strong interactions. 
In Figure 2 we show a model realizing this non-trivial 2D state. The model is
 constructed out of orthogonally crossed 1D FK chains, and each unit cell contains 16 complex fermions $\psi^J_{\mb{n}}$, $J= 1,\dots,16$,
where $\mb{n}$ now labels a site on the square lattice. This model exhibits a non-trivial weak invariant 
$\mb{G}_{\nu}=\frac{1}{2}(4\mb{b}_1+4\mb{b}_2)$, and is a 2D topological charge-$4e$ superconductor, using the same interpretation discussed for the 1D model. We note that an almost identical discussion could be had for $C_2$ rotation or a reflection symmetry with, for example, $\mb{G}_{\nu}=\tfrac{1}{2}(4\mb{b}_1)$ or $\tfrac{1}{2}(4\mb{b}_2).$

\begin{figure}[t]
  \centering
    \includegraphics[width= .5\textwidth]{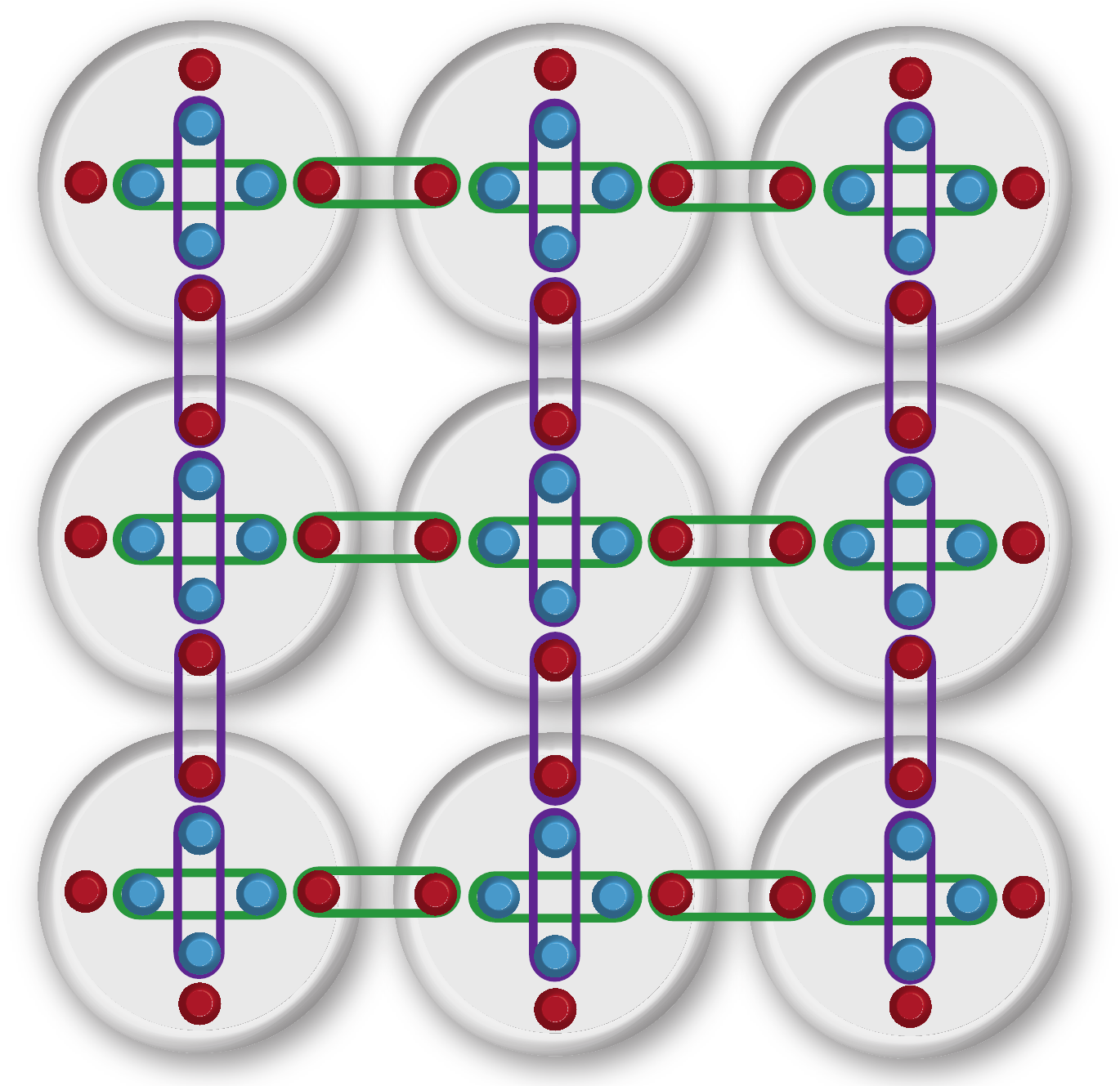} 
\vskip 10pt
 \caption{\emph{Topological Crystalline Superconductor in 2D:} A 2D model of a time-reversal invariant, $C_4$-symmetric  topological crystalline superconductor
 with four unpaired Majorana fermions in each boundary unit cell. This model is made of crossed vertical and horizontal FK chains, 
and so each
unit cell (large white circles) contains 16 complex (32 Majorana) fermions. To reduce clutter in the figure, each red circle represents four a-type Majorana
fermions and each blue circle represents four b-type Majorana fermions. The green lines indicate a FK interaction in the
horizontal wires and the purple lines indicate a FK interaction in the vertical wires.}
\label{fig:2Dmodel}
\end{figure}

A non-trivial $\mb{G}_{\nu}$ of this form reflects the fact that on a system with open boundary conditions along an edge with a normal vector parallel to $\mb{b}_1$ or $\mb{b}_2$,  each boundary unit cell will contain four unpaired Majorana modes of the same type. As discussed earlier in the context of the 1D FK chain, the Hilbert
space for these four Majorana modes consists of two spin 1/2 degrees of freedom per boundary site, and at least
one of these spin 1/2's always remains gapless in the presence of generic edge perturbations. This is a consequence of the fact that the time-reversal operator acts projectively as $T^2 = -1$ on these boundary spin 1/2 degrees of freedom, whereas it acts as $T^2 = 1$
on the complex fermions $\psi^J_{\mb{n}}$ in each unit cell. Generically then, this model has a translation-symmetric, anti-ferromagnetic Heisenberg spin-1/2 chain on its edge. This non-chiral, gapless boundary has an identical low energy description to the critical 1D FK chain discussed above, and it is protected by translation and time-reversal symmetry. The fact that the system is electronic implies that the boundary is anomalous, i.e., it cannot be realized in a pure 1D electronic lattice system with the same symmetries. Recall that an electron operator $\psi=(a+ib)/2$ is a combination of both $a$- and $b$-type Majorana fermions. If the boundary modes could be gapped out by coupling them to an external 1D wire of electrons with the same symmetries, then the edge would not be anomalous. Here we see that this is not the case as the external wire would contain both $a$- and $b$-type fermions to preserve $T.$ The free $a$-type tetrads on the edge could be annihilated by coupling to the external $b$-type fermions (through tunneling terms) or the external $a$-type tetrad (via the FK interaction), however in either case this would leave four unpaired Majorana fermions on the external wire in the low-energy sector unless one of the symmetries is broken. Thus, modifying the edge can convert $a$-type Majorana fermions to $b$-type, but always leaves a set of gapless modes. 

In addition to non-trivial boundary states, the crystalline symmetry-protected topology gives rise to topological qubits (i.e., non-Abelian excitations) at (semiclassical) lattice defects. Based on the work of Refs.~\citenum{RanZhangVishwanath,TeoKane,teohughes1,teohughes2} we can determine by inspection that a dislocation with Burgers vector $\mb{B}$ will have $\tfrac{1}{\pi}\mb{B}\cdot\mb{G}_{\nu}$ Majorana bound states at the core. Additionally, a vertex-type disclination with Frank angle $\Omega=\pm\pi/2$ will also trap a tetrad of unpaired $a$-type Majorana bound states, while a plaquette-type disclination with $\Omega=\pm\pi/2$ will not trap any unpaired Majorana modes.
Each non-trivial defect $\Sigma$, of either kind, thus binds a decoupled tetrad of $a$-type Majorana fermions. Adding the local quartic perturbation (c.f. $H_{bdy}$) reduces the Majorana tetrad to a single spin-$1/2$ degree of freedom which is identical to an end of the topological FK-chain. Therefore each of these defects carries a quantum dimension $d_\Sigma=2$, which signifies their non-Abelian nature and ability to store quantum information non-locally in space. We comment more on this in Appendix C where we also discuss the fusion rules of these defects. We note that since these defects are extrinsic/semiclassical, their projective braiding properties can be determined, but we leave this for future work. 

\paragraph*{Conclusion and more discussions.} 
There has been an exciting discussion of a charge $4e$ superconductor in another recent work by Berg et al.~in Ref.~\citenum{BergFradkinKivelson}, and follow up work by other researchers in Refs.~\citenum{herland2010,moon2012skyrmions} (charge $6e$ superconductors
were considered in Ref.~\citenum{agterberg2011}).
 Let us briefly discuss the similarities and distinctions between our model and the charge $4e$ superconductor considered by
Berg et al. in \citenum{BergFradkinKivelson}. In both systems the tetrad $\langle\psi_a\psi_b\psi_c\psi_d\rangle$ is the only quasi-long range order. As the ground state consists of $4e$ bosons, Cooper pairs are actually {\em fractional} excitations that require finite energy, and cannot ``disappear" into the condensate. Magnetic flux is quantized now in units of $hc/4e$ and if such a material exists, this in principle could be measured by SQUID loops or Josephson junctions. The scenario to generate the $4e$ superconductivity in \citenum{BergFradkinKivelson} was based on a striped superconductor with a unidirectional incommensurate pair density wave. Their $4e$ superconductor is a melted phase with restored continuous translation symmetry. Our model on the other hand is a strong-coupling type construction with discrete lattice symmetries that arises as an effective description of (possibly) an eight-body interacting Hamiltonian when the charge conservation symmetry is broken spontaneously. Strong interactions enable the non-trivial topology -- which is absent in \citenum{BergFradkinKivelson} -- to coexist with crystalline symmetries. Unlike a striped superconductor, we see that dislocations and disclinations in our two-dimensional model trap topologically protected non-Abelian modes. While the dislocations in the Berg et al.~state bind vortices, the vortices do not carry protected non-Abelian excitations, and are Abelian defects.

In conclusion, we have shown that interactions can allow for a general mechanism to produce interacting topological phases that have no free-fermion description. For the cases considered here, we found topological, charge-$4e$ superconductors which were protected by inversion and rotation symmetries. The boundaries of these systems, and their topological defects, trap low-energy degrees of freedom which could be used for the robust, non-local storage of quantum information.

We wish to acknowledge T. Morimoto for useful conversations. TLH is supported by ONR award N0014-12-1-0935. JCYT 
acknowledges support from the Simons Foundation. We thank the ICMT at UIUC for support.

\appendix
\section{Hamiltonian for the lattice model and critical theory} 

Using the idea of Fidkowski and Kitaev\cite{FK2010}, any eight Majorana fermions $a^1,\ldots,a^8$ of the same type can be gapped out by a time-reversal symmetric four-body Hamiltonian
\begin{widetext}
\begin{align}H_{FK}=-u\left[a^{1234}+a^{5678}+\sum_{\sigma\in A_4}a^{\sigma(1)\sigma(2)[\sigma(1)+4][\sigma(2)+4]}\left(\frac{1+a^{1234}}{2}\right)\right]\label{FKham}\end{align} 
\end{widetext}
where $a^{IJKL} \equiv a^Ia^Ja^Ka^L$ and $\sigma$ runs over even permutations of $(1234)$ (i.e. the 12 elements of the
alternating group $A_4$). For each green line connecting a set of eight Majorana fermions in Figures 1b,c and 2 we have a copy of $H_{FK}.$ From these figures we see that the $a$-type and $b$-type Majorana fermions in the FK-chain are coupled separately, i.e., the eight $b$-fermions are gapped by an intra-cell FK interaction, while the $a$-fermions couple in-between unit cells similar to special limits of the familiar Kitaev p-wave wire model\cite{kitaev2001}. 

We can understand $H_{FK}$ in the following way. Each quartet of Majorana fermions represents two spin-1/2 degrees of freedom, one with even fermion parity and one with odd (see below). The terms $-u(a^{1234}_n  + a^{5678}_n)$ lock the local fermion parity to $+1$ (if $u>0$) for both quartets, leaving only a single spin-1/2 effective degree of freedom corresponding to each quartet.  Now, using a Jordan-Wigner transformation, the remaining low-energy degrees of freedom in the FK-chain can be transformed to the antiferromagnetic Heisenberg spin-$1/2$ chain with two spins per unit cell (since there are two $a$-type quartets per unit cell, and the two $b$-type quartets are gapped by the intra-cell interaction). The topological FK-chain, with locally frozen fermion parity, therefore shares some properties of the Haldane spin chain and supports topological spin-$1/2$ excitations at its ends. The topological phase transition is driven by turning on intra-cell FK couplings for the $a$-type Majorana fermions. As mentioned above, the critical theory -- when the intra-cell interaction strength matches that of the inter-cell one -- can be described by a conformal field theory with central charge $c=1$. It can be mapped into a non-chiral $u(1)_1$ boson theory with compactification radius $R=\sqrt{2}$ and carries an affine Kac-Moody $su(2)$ structure at level 1. 

%\cite{Afflecklecture}

\section{Proof that 1D boundary states have $T^2=-1$}

The complex fermion operators $\chi_1 = \frac{1}{2}(a^1 + ia^2)$, $\chi_2 = \frac{1}{2}(a^3 + ia^4)$, which are used in the definition of $H_{bdy}$, obey the unconventional transformation $T\chi_j T^{-1} = \chi^{\dg}_j$. Now, suppose
$|\tilde{0}\ran$ is the state annihilated by $\chi_1$ and $\chi_2$. Then $H_{bdy}$, with $\lambda>0,$ has the two ground states: 
$|\tilde{0}\ran$ and $|\tilde{1}\ran \equiv \chi^{\dg}_1\chi^{\dg}_2|\tilde{0}\ran$. These states transform
non-trivially under the action of $T$. To see this, first note that since $T$ is anti-unitary,  we have 
$\lan T \Psi | T \Psi \ran = \lan \Psi | \Psi \ran^\ast$ for any state $|\Psi\ran.$ In particular this means that $T| \Psi\ran \neq 0$ if $\lan \Psi |\Psi\ran \neq 0$.
We see then that 
$T |\tilde{1}\ran$ can only be non-zero if $T  |\tilde{0}\ran \propto |\tilde{1}\ran$.
We can choose the convention $T  |\tilde{0}\ran = |\tilde{1}\ran$, which just amounts to a 
choice of phase since $\chi^{\dg}_1$ and $\chi^{\dg}_2$ anti-commute. Using
this rule we also find that $T |\tilde{1}\ran = - |\tilde{0}\ran$.
So in the basis of ground states on the edge, $|\tilde{0}\ran$ and $|\tilde{1}\ran$, 
the time-reversal operator acts non-trivially as a matrix $T_{bdy} = i\sigma^y K$ 
(where $K$ is complex conjugation in this basis),
such that $(T_{bdy})^2 = -1$ at the edge of 
our system. It follows immediately from Kramer's theorem that the remaining double degeneracy of the boundary states is protected against 
arbitrary perturbations that do not break time-reversal symmetry.
Thus, we see that the low energy degrees of freedom on the edge form a projective representation of the on-site $\mathbb{Z}_2$ symmetry group 
generated by  $T$ ($T^2 = 1$) on the local degrees of freedom in the bulk of the system. This represents an alternate proof that 
$T^2 = -1$ on the edge of the $\nu=4$ phase of the one-dimensional Majorana chain of class BDI, which was also shown in Refs.~\citenum{FK2011,TPB2011,Wen2011complete}.

\section{Properties of Defects} 

Consider two non-trivial defects (say a pair of dislocations) in the 2D TCS. On each defect the stable degree of freedom is a single spin-1/2. Analogous to the tensor product of a pair of spins $[\tfrac{1}{2}]\otimes[\tfrac{1}{2}]=[0]\oplus[1]$, a pair of separated defects $\Sigma_1, \Sigma_2$ is associated to a fourfold degeneracy seen by the defect fusion \begin{align}\Sigma_1\times\Sigma_2=1+\psi_1\psi_2+\psi_1\psi_3+\psi_1\psi_4\ .\end{align}
 The vacuum channel $1$ is the ground state if the dislocation pair is coupled by the FK interaction. It corresponds to the singlet channel for the pair of spins. The other three are time-reversal breaking ground states when the dislocations are coupled by Eq. \eqref{FKham} but with a reversed sign in front of the sum over even permutations (the sum over $\sigma\in A_4$). This corresponds to the ferromagnetic Heisenberg interaction and the three states are the tensor products $|\uparrow\uparrow\rangle_X,|\uparrow\uparrow\rangle_Y,|\uparrow\uparrow\rangle_Z$ with respect to spin-up in the $x,y,z$ directions. %Unlike the triplet channels of a pair of spins, they do not group into a single sector unless the local $SO(4)$ symmetry is gauged. 
It is convenient to choose these non-orthogonal, but still linearly independent, basis vectors. For example the $\psi_1\psi_2$ channel has a non-trivial vacuum expectation value for the Cooper pair $\langle\psi_1\psi_2\rangle=-\langle\psi_3\psi_4\rangle=i$. In a defect-less $4e$ superconductor, Cooper pairs are gapped excitations and are not responsible for transport at low temperatures. The presence of these non-Abelian defects could provide {\em wormholes} for Cooper pairs to teleport. A non-vanishing charge $2e$ tunneling between two normal BCS superconductor leads in contact with a $4e$ superconductor would therefore be a signature for the non-trivial topology.

%\bibliography{InteractingWeakBib}

\end{document}